\documentclass[12pt]{iopart}

\usepackage{graphics}
\begin{document}

\title[QCD plasma thermalization, collective flow and extraction of shear
viscosity]
{QCD plasma thermalization, collective flow and extraction of shear viscosity
\footnote{Invited talk.}}

\author{Zhe Xu$^1$, Carsten Greiner$^1$ and Horst St\"ocker$^{1,2}$}

\address{$^1$ Institut f\"ur Theoretische Physik,
Goethe-Universit\"at Frankfurt, Max-von-Laue-Str.1, D-60438,
Frankfurt, Germany\\
$^2$Gesellschaft f\"ur Schwerionenforschung mbH (GSI),
Planckstr. 1, D-64291, Darmstadt, Germany}
\ead{xu@th.physik.uni-frankfurt.de}
\begin{abstract}
Fast thermalization and elliptic flow of QCD matter found at the
Relativistic Heavy Ion Collider (RHIC) are understood as the consequence
of perturbative QCD (pQCD) interactions within a (3+1) dimensional parton
cascade. The main contributions stem from pQCD-inspired bremsstrahlung.
We extract the shear viscosity to entropy ratio, which is between
$0.08$ and $0.15$.
\end{abstract}

\maketitle

\section{Introduction and motivation}
The values of the elliptic flow parameter $v_2$ measured by the experiments at
the Relativistic Heavy Ion Collider (RHIC) \cite{rhicv2} are so large as
predicted by calculations employing the ideal hydrodynamics \cite{H01}.
This finding suggests at first that a fast local equilibration of quarks
and gluons occurs at a very short time scale, $\le 1$ fm/c, as used in the
hydrodynamical calculations. Coherent quantum effects like color
instabilities \cite{instab} may play a role in isotropization of particle
degrees of freedom at the very initial stage where the matter is super dense.
However, more quantitative studies are needed to determine their significance
on the true thermal equilibration as suggested for the expanding quark-gluon
matter at RHIC.

The second message stemming from a comparison between measured $v_2$ and
that predicted by calculations is that the locally thermalized state of
matter created, the quark-gluon plasma (QGP), behaves as a nearly perfect
fluid exhibiting strong ``explosive'' collective motion. Quarks and gluons
should be strongly coupled \cite{sqgp}. The reason for it is still an open
issue. Furthermore, recent investigations \cite{L07} show that the QGP created
should indeed possess a small viscosity coefficient. New viscous
hydrodynamical calculations \cite{RR07} are worked out to extract
the viscosity from comparisons to experimental data. Drawbacks in these
calculations are the assumed ideal thermal initial conditions and
the assumed Bjorken longitudinal expansion. The smallness of the viscosity
is of great interest as a result of the recent debate about speculative
``realizations'' of super symmetric representations of Yang-Mills theories
using the AdS/CFT conjecture \cite{adscft}. However, before one may
believe that an introduction of extra effective degrees of freedom is
really needed, one should also pursue more conservative explanations for
strongly coupled system, using well-established theories. In this talk,
we demonstrate that the perturbative QCD (pQCD) can still explain a fast
thermalization of the initially nonthermal gluon system, the large collective
effects of QGP created at RHIC and the smallness of the shear viscosity
to entropy ratio in a consistent manner by using a relativistic pQCD-based
on-shell parton cascade Boltzmann approach of multiparton scatterings
(BAMPS) \cite{XG05,XG07,XG08,XGS08}. In principle, there is no need to
invoke exotic black hole physics in higher dimensions and super symmetric
Yang-Mills theories using the AdS/CFT conjecture to understand RHIC data.

\section{Parton cascade BAMPS}
BAMPS is a parton cascade, which solves the Boltzmann transport equation
and can be applied to study, on a semi-classical level, the dynamics of
gluon matter produced in heavy-ion collisions at RHIC energies. 
The structure of BAMPS is based on
the stochastic interpretation of the transition rate \cite{XG05,DB91,L93,C02},
which ensures full detailed balance for multiple scatterings. BAMPS subdivides
space into small cell units where the operations for transitions are performed.

Gluon interactions included in BAMPS are elastic pQCD $gg\to gg$ 
scatterings as well as pQCD-inspired bremsstrahlung
$gg\leftrightarrow ggg$. The differential cross sections and the effective
matrix elements are given by \cite{biro}
\begin{eqnarray}
\label{cs22}
\frac{d\sigma^{gg\to gg}}{dq_{\perp}^2} &=&
\frac{9\pi\alpha_s^2}{(q_{\perp}^2+m_D^2)^2}\,,\\
\label{m23}
| {\cal M}_{gg \to ggg} |^2 &=&\frac{9 g^4}{2}
\frac{s^2}{({\bf q}_{\perp}^2+m_D^2)^2}\,
 \frac{12 g^2 {\bf q}_{\perp}^2}
{{\bf k}_{\perp}^2 [({\bf k}_{\perp}-{\bf q}_{\perp})^2+m_D^2]}
\Theta(k_{\perp}\Lambda_g-\cosh y)
\end{eqnarray}
where $g^2=4\pi\alpha_s$. ${\bf q}_{\perp}$ and
${\bf k}_{\perp}$ denote the perpendicular component of the momentum
transfer and of the radiated gluon momentum in the center-of-mass
frame of the collision, respectively. $y$ is the momentum rapidity of
the radiated gluon in the center-of-mass frame, and $\Lambda_g$ is the
gluon total mean free path, which is calculated self-consistently \cite{XG05}.
The interactions of the massless gluons are screened by a Debye mass
$m_D^2=\pi d_G \,\alpha_s N_c \int d^3p /(2\pi)^3 \cdot f / p$
where $d_G=16$ is the gluon degeneracy factor for $N_c=3$.
$m_D$ is calculated locally using the gluon density function $f$
obtained from the BAMPS simulation. The suppression of the bremsstrahlung
due to the Landau-Pomeranchuk-Migdal (LPM) effect is taken into account
within the Bethe-Heitler regime employing the step function
in equation (\ref{m23}). 

The initial gluon distributions are taken as an ensemble of minijets with
transverse momenta greater than $1.4$ GeV \cite{XG07}, produced via
semihard nucleon-nucleon collisions. We use Glauber geometry with
Woods-Saxon profile and assume independent binary nucleon-nucleon
collisions. A formation time for initial minijets is also
included \cite{XG05,XG07}.

In the present simulations, the interactions of the gluons are
terminated when the local energy density drops below $1\ \rm{GeV fm}^{-3}$.
This value is assumed to be the critical value for the occurrence of 
hadronization, below which parton dynamics is not valid. Because
hadronization and then hadronic cascade are not yet included in BAMPS,
a gluon, which ceases to interact, propagates freely and can be
regarded as a free pion employing a picture of parton-hadron duality.
Implementing a Cooper-Frye prescription for hadronization and employing
UrQMD \cite{urqmd} for the hadronic cascade are in progress.

The minijet initial conditions and the subsequent evolution using
the present prescription of BAMPS for two sets of the coupling
$\alpha_s=0.3$ and $0.6$ give nice agreements to the measured 
transverse energy per rapidity over all rapidities
 \cite{XGS08,phobos-brahms}.

\section{Thermalization of gluon matter}
To study possible thermalization of gluons we concentrate on the local
central region which is taken as an expanding cylinder with a radius
of $1.5$ fm and within an interval of space time rapidity
$-0.2 < \eta < 0.2$, where $\eta=\frac{1}{2}\ln[(t+z)/(t-z)]$.
Figure 1 shows the varying transverse momentum
spectrum with time obtained from the BAMPS calculations for central Au+Au
collisions at $\sqrt{s}=200$A GeV, with elastic pQCD $gg\to gg$ only
(left panel) and including pQCD-inspired bremsstrahlung
$gg \leftrightarrow ggg$ (right panel), respectively.
\begin{figure}[h]
\label{dndpt}
\begin{center}
\resizebox{0.44\textwidth}{!}{
  \includegraphics{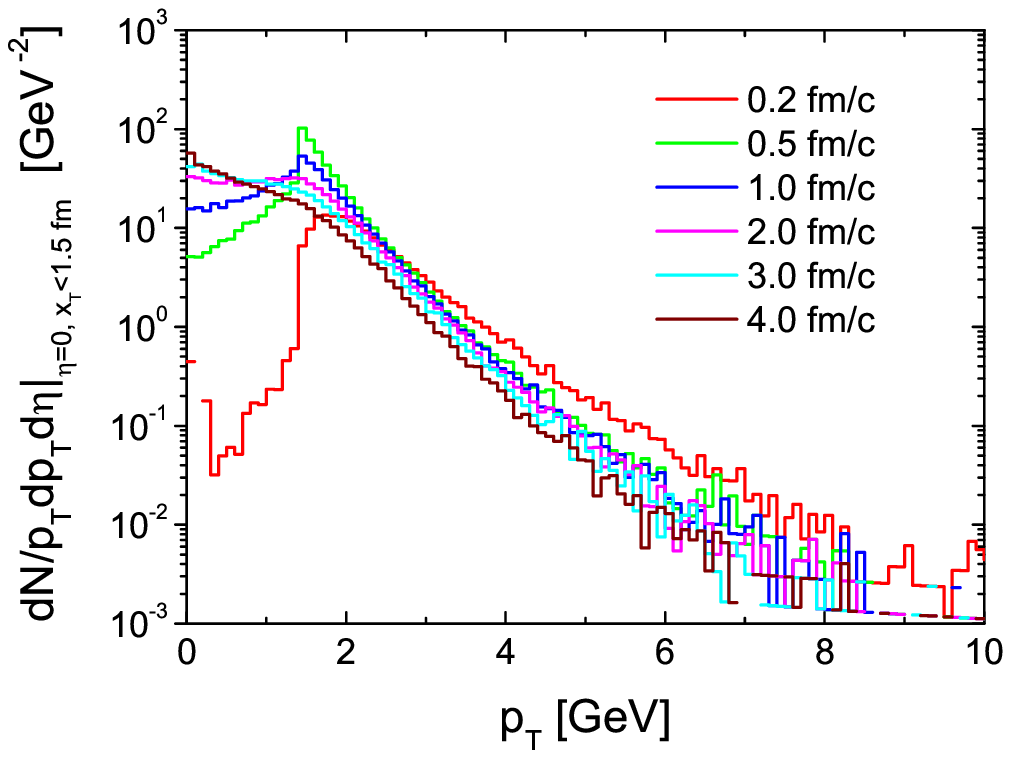}
}
\hspace{\fill}
\resizebox{0.44\textwidth}{!}{
  \includegraphics{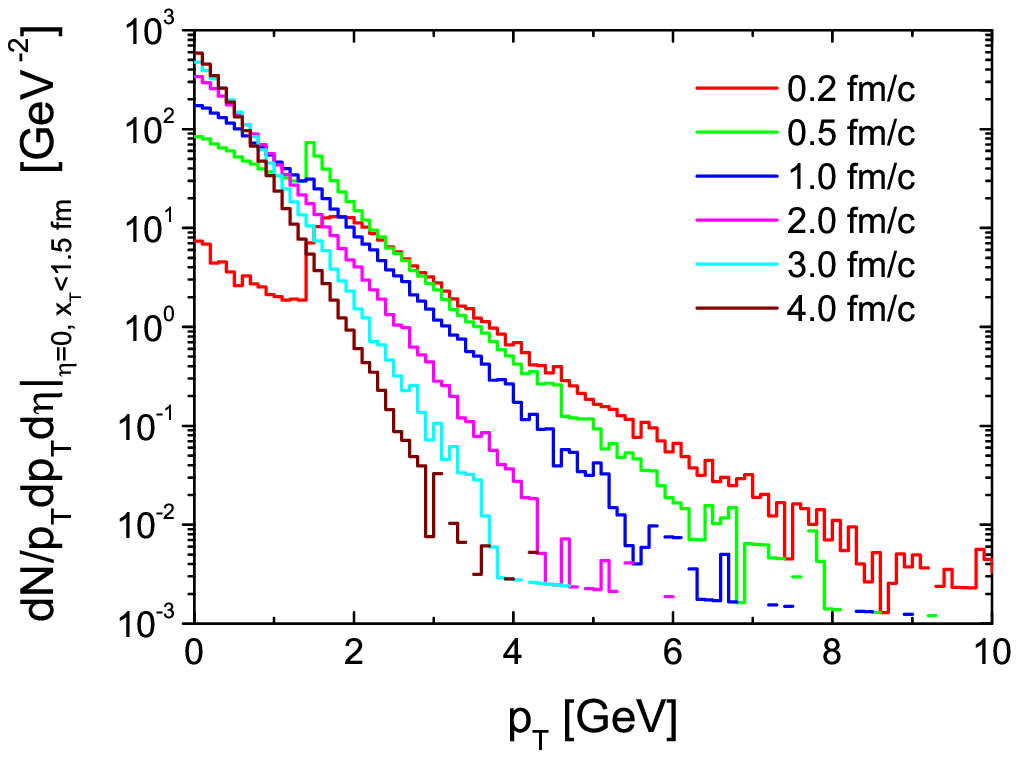}
}
\caption{Transverse momentum spectrum in the central region at
different times obtained from the BAMPS simulation
with $gg \to gg$ only (left panel) and including 
$gg \leftrightarrow ggg$ collisions (right panel).
}
\end{center}
\end{figure}
$\alpha_s=0.3$ is used. One clearly sees that not much happens in the
left panel of figure 1. With only elastic pQCD interactions the gluon
system is initially nonthermal and also stays in a nonthermal state
at late times. The situation is dramatically changed when the inelastic
interactions are included. In the right panel of figure 1 the curves from
the upper to lowest depict, respectively, the spectrum at $t=0.2$, $0.5$,
$1$, $2$, $3$, and $4$ fm/c. We see that the spectrum reaches an
exponential shape at $1$ fm/c and becomes increasingly steeper at late
times. This is a clear indication for the achievement of local thermal
equilibrium and the onset of hydrodynamical collective expansion with
subsequent cooling by longitudinal work.

The inelastic pQCD-based bremsstrahlung and its back reaction are
essential for the achievement of local thermal equilibrium at a short
time scale. The fast thermalization happens in a similar way if color
glass condensate is chosen as the initial conditions \cite{EXG08}.
One of the important messages obtained there is that the hard gluons
thermalize at the same time as the soft ones due to the $ggg\to gg$
process, which is not included in the ``Bottom-Up'' scenario of
thermalization \cite{BMSS01}.

\section{Significance of $gg\leftrightarrow ggg$ in thermalization
and its crucial role in viscosity}
To understand the efficiency of the pQCD $gg\leftrightarrow ggg$
processes for thermalization, we first calculate the collision rates
$R_i=n\langle v_{\rm{rel}} \sigma_i \rangle$, $i=gg\to gg, gg\to ggg$,
using equations (\ref{cs22}) and (\ref{m23}) in a thermal gluon gas.
We find that $R_{gg\to gg}$ is larger than $R_{gg\to ggg}$, which
implies that the collision rate is not the correct quantity determining 
the contribution of different processes to thermalization. Moreover,
for instance, $\langle v_{\rm rel} \sigma_{gg\to gg} \rangle=0.82$ mb
and $\langle v_{\rm rel} \sigma_{gg\to ggg} \rangle=0.57$ mb at
temperature $T=400$ GeV and $\alpha_s=0.3$. A small value of cross
sections can still lead to a fast equilibration.

Kinetic equilibration relates to momentum deflection. Large momentum
deflections due to large-angle scatterings will speed up kinetic
equilibration enormously. Whereas the elastic pQCD scatterings favor
small-angle collisions, the collision angles in bremsstrahlung
processes are almost isotropically distributed at RHIC energy due to
the incorporation of the LPM effect \cite{XG05,XG07}. This is the
intuitive reason why the bremsstrahlung processes are more effective
in equilibration than the elastic interactions.

Quantitatively we demonstrated in \cite{XG07} that the
contributions of different processes to momentum isotropization are
quantified by the so-called transport rates
\begin{equation}
\label{trate}
R^{\rm tr}_i= \frac{\int \frac{d^3p}{(2\pi)^3} \frac{p_z^2}{E^2} C_i -
\langle \frac{p_z^2}{E^2} \rangle \int \frac{d^3p}{(2\pi)^3} C_i}{n\,
(\frac{1}{3}- \langle \frac{p_z^2}{E^2} \rangle)} \,,
\end{equation}
where $C_i[f]$, functional of the gluon density distribution $f(p,x)$,
is the corresponding collision term describing various interactions,
$i=gg\to gg, gg\to ggg, ggg\to gg$, respectively.
Their sum gives the inverse of the time scale of momentum
isotropization, which also marks the time scale of overall thermalization.
To obtain $R^{\rm tr}_i$ close to thermal equilibrium we take
$f=e^{-\beta E} \left (1-\frac{\chi}{2}\, \beta \,\frac{p_z^2}{E} \right )$
with a small $\chi$ and calculate the term in zeroth order of $\chi$ in
equation (\ref{trate}) \cite{XG08}. Figure 2 shows the transport collision
rate, scaled by temperature $T=1/\beta$, for elastic $gg\to gg$ scattering
and bremsstrahlung $gg\to ggg$, respectively.
\begin{figure}[ht]
\label{rates}
\begin{center}
\resizebox{0.6\textwidth}{!}{
  \includegraphics{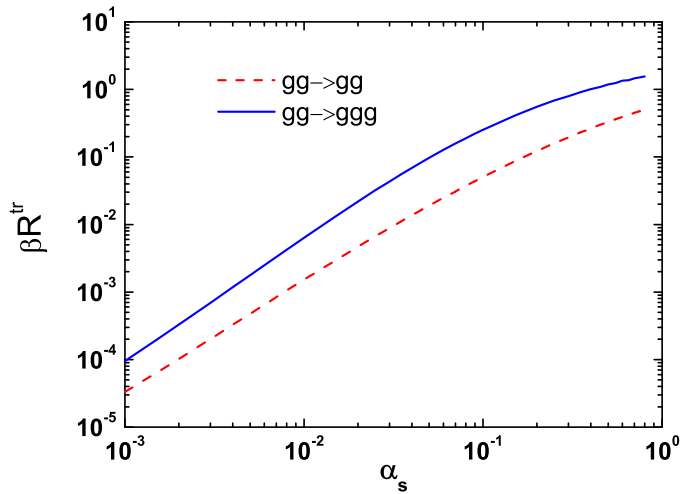}
}
\caption{Transport collision rates scaled by temperature.
}
\end{center}
\end{figure}
$R^{\rm tr}_{gg\to ggg}$ is a factor of $3-5$ larger than
$R^{\rm tr}_{gg\to gg}$ over a range in the coupling constant $\alpha_s$
from $10^{-3}$ to $0.8$, which demonstrates the essential role of the
bremsstrahlung in thermal equilibration. Because the transport rate
(\ref{trate}) is expressed by characteristic moments of the collision
term, it certainly contains a distribution of the collision angles,
although it is not directly visible as in the transport cross section 
$\sigma^{\rm tr}_i=\int d\sigma_i \sin^2\theta$.

For a gluon gas, which is initially far away from equilibrium, we can
roughly estimate the time scale of thermalization $\tau_{\rm eq}$ by
taking the inverse of the sum of the transport collision rates close to
thermal equilibrium. At temperature $T=400$ MeV we obtain
$\tau_{\rm eq}\approx 1/\sum R^{\rm tr}=0.32$ fm/c for $\alpha_s=0.3$.
We note that the above hinges on the assumption that the system is static.
Expanding systems are more complicated because particles flow, which
drives the systems out of local equilibrium. Therefore, the momentum
degradation of flowing particles toward isotropy is slower than the
inverse of the total transport collision rate \cite{XG07}.

Using the Navier-Stokes approximation we have derived the shear
viscosity $\eta$ from relativistic kinetic theory and found \cite{XG08}
\begin{equation}
\label{shv}
\eta \cong \frac{1}{5} n \frac{\langle E(\frac{1}{3}-\frac{p_z^2}{E^2}) \rangle}
{\frac{1}{3}-\langle \frac{p_z^2}{E^2} \rangle} \frac{1}{\sum R^{\rm tr}+ 
\frac{3}{4} n \partial_t (\ln \lambda)}\,,
\end{equation}
where $\lambda$ denotes the gluon fugacity. This expression gives
a direct correspondence of the macroscopic quantity $\eta$ to its
microscopic origin: $\eta$ is inversely proportional to the sum of
the total transport collision rate and the chemical equilibration rate,
and it is roughly proportional to the energy density $\epsilon$. At thermal
equilibrium we obtain
\begin{equation}
\label{shv2}
\eta=\frac{4}{15} \, \frac{\epsilon}{\sum R^{\rm tr}}
\end{equation}
and the shear viscosity to entropy ratio is
\begin{equation}
\label{etas}
\frac{\eta}{s}=\left ( 5 \beta \sum R^{\rm tr} \right )^{-1}=
\left (5 \beta R^{\rm tr}_{gg \to gg}+ 
\frac{25}{3} \beta R^{\rm tr}_{gg \to ggg} \right )^{-1}
\end{equation}
where the entropy density $s=\frac{4}{3} \beta \epsilon$ is used.
Within the present description bremsstrahlung and its back reaction
lower the shear viscosity to entropy density ratio significantly by
a factor of $7$, compared with the ratio when only elastic collisions
are considered. For $\alpha_s=0.3$ we obtain $\eta/s=0.13$. To match
the lower bound of $\eta/s=1/4\pi$ from the AdS/CFT
conjecture \cite{adscft} $\alpha_s=0.6$ has to be chosen.
Even for that case the cross sections are in the order of $1$ mb for
a temperature of $400$ MeV. We see that perturbative QCD interactions
can drive the gluon matter to a strongly coupled system with an
$\eta/s$ ratio as small as the lower bound from the AdS/CFT conjecture.

\section{Collective flow $v_2$ and extraction of shear viscosity}
The elliptic flow $v_2=\langle (p_x^2-p_y^2)/p_T^2 \rangle$ can be
directly calculated from microscopic simulations for Au+Au collisions
at $\sqrt{s}=200$A GeV employing BAMPS. $\alpha_s=0.3$ and $0.6$ are
used for comparisons. For a firm footing we compare our results with
the experimental data, assuming parton-hadron duality. 
The left panel of figure 3 shows the elliptic flow $v_2$ at midrapidity
for various centralities (impact parameters), compared with the
PHOBOS \cite{phobos} and STAR \cite{star} data.
\begin{figure}[h]
\label{v2shv}
\begin{center}
\resizebox{0.44\textwidth}{!}{
  \includegraphics{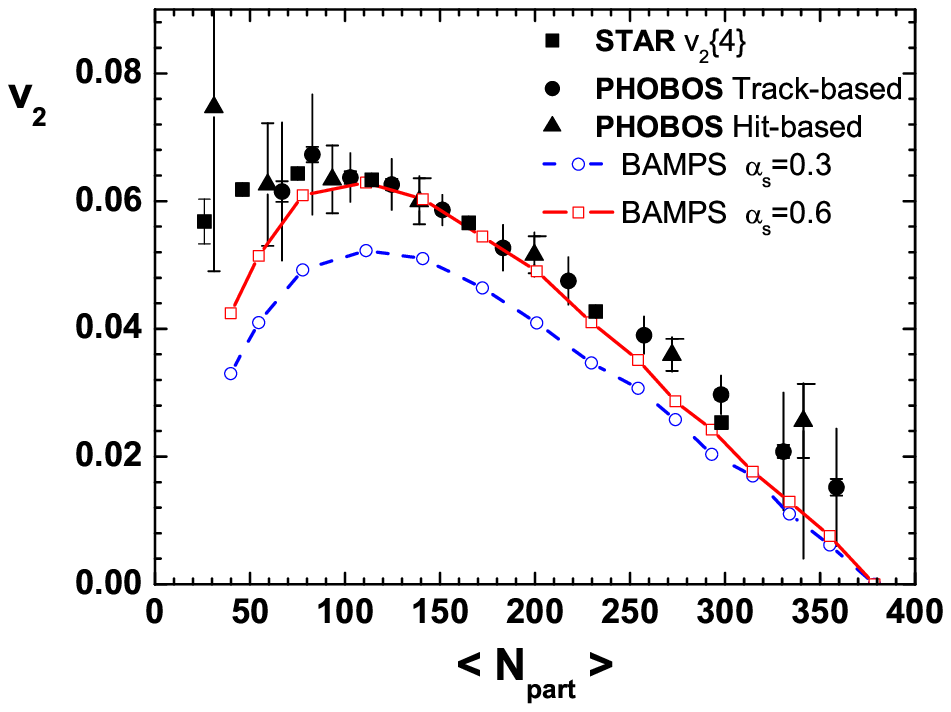}
}
\hspace{\fill}
\resizebox{0.44\textwidth}{!}{
  \includegraphics{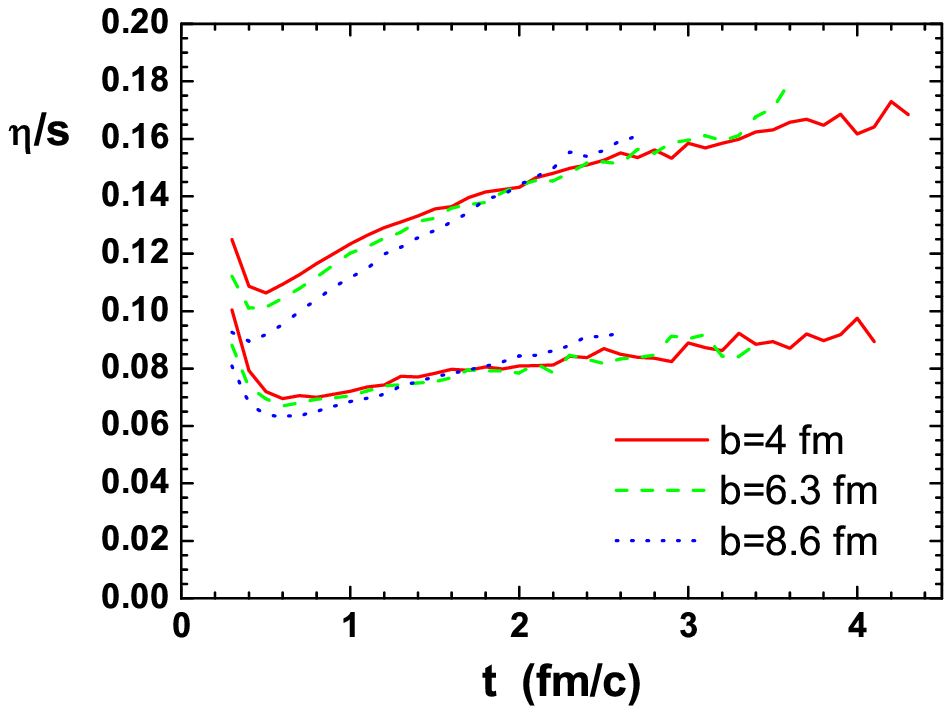}
}
\caption{Left panel: elliptic flow $v_2(|y|<1)$ from BAMPS
using $\alpha_s=0.3$ and $0.6$, compared with the PHOBOS \cite{phobos}
and STAR \cite{star} data. Right panel: the shear viscosity to entropy density
ratio $\eta/s$ at the central region during the entire expansion. $\eta/s$
values are extracted from the simulations at impact parameter $b=4$,
$6.3$, and $8.6$ fm. The upper band shows the results with $\alpha_s=0.3$
and the lower band the results with $\alpha_s=0.6$.
}
\end{center}
\end{figure}
Except for the central centrality region the results with $\alpha_s=0.6$
agree perfectly with the experimental data, whereas the results with
$\alpha_s=0.3$ are roughly $20\%$ smaller. We see that the generation of
the large elliptic flow observed at RHIC is well described by pure
perturbative gluon interactions as incorporated in BAMPS.

To see how viscous the gluon plasma behaves, the ratio of the shear
viscosity to the entropy density is extracted by using equation (\ref{shv})
in those spatial regions where the matter is nearly equilibrated and,
thus, behaves quasi-hydrodynamically, i.e., like a viscous fluid.
The entropy density is estimated by
$s=4n-n\,\ln \lambda$ assuming that the gluon matter is in local
kinetic equilibrium. The true entropy density is expected to be
(slightly) smaller than that calculated by the above formula, because
overall kinetic equilibration cannot be complete in an expanding system.
Thus, the true shear viscosity to entropy density ratio is (slightly)
larger than that calculated.

The ratio of the shear viscosity to the entropy density, $\eta/s$, is
shown in the right panel of figure 3. We see that the ratio
does not depend strongly on gluon density or temperature, since
interaction rates and transport collision rates scale with the temperature.
Hence, $\eta/s$ depends practically only on $\alpha_s$. For $\alpha_s=0.6$,
at which the $v_2$ values match the experimental data, we obtain
$\eta/s \approx 0.08$, which is the same as the lower bound from the AdS/CFT
conjecture. However, $\eta/s$ may be higher, since the inclusion of
hadronization \cite{MV03} and subsequent hadronic cascade \cite{ZBS05}
will yield, even only moderate, contributions to the final elliptic flow
values. Furthermore, different picture of initial conditions (e.g. color
glass condensate) will also lead to different initially spatial eccentricity
and, hence, will affect the final value of $v_2$ and the result on $\eta/s$.
These investigations are underway and will provide more constraints on
extracting $\eta/s$.

Note that the relation between $\eta/s$ and $v_2$ (as a function of 
$\langle N_{\rm part} \rangle$) is amazingly consistent with recent
findings from a special set of viscous hydrodynamical
calculations \cite{RR07}. This certainly deserves further investigations.

\section{Conclusion}
The pQCD-based parton cascade BAMPS is used to calculate the time scale 
of thermalization, the elliptic flow $v_2$ and to extract the ratio of
the shear viscosity to the entropy density, $\eta/s$, from simulations
of Au+Au collisions at RHIC energy $\sqrt{s}=200$A GeV. This is a
committed and large-scale undertaking. The present BAMPS includes
elastic $gg\to gg$ and inelastic bremsstrahlung and its back reaction
$gg \leftrightarrow ggg$. We observed that the gluon matter thermalizes at
$1$ fm/c for central collisions. Agreement with the experimental data
on $v_2$ is found with Glauber-type minijets initial conditions and
$\alpha_s=0.3-0.6$. The $\eta/s$ ratio of the gluon plasma created
varies between $0.08$ and $0.15$. Standard pQCD interactions alone can
describe the generation of large $v_2$ values at RHIC. The small $\eta/s$
ratios found in the simulations indicate that the gluon plasma created behaves
like a nearly perfect fluid. This can be understood by perturbative QCD
without resorting to exotic explanations such as the AdS/CFT conjecture.
Gluon bremsstrahlung dominates and yields rapid thermalization, and,
therefore, early pressure buildup, and a small shear viscosity in the
gluon gas. Many further analyses on jet quenching and quark degrees of
freedom including hadronization are underway to establish a more 
global picture of heavy-ion collisions.

\end{document}